\documentclass[twocolumn,floatfix,showpacs,amsmath,amssymb,prb]{revtex4}
\usepackage{graphicx}
\begin{document}

\title{The ferromagnetic transition and domain structure in LiHoF$_4$ }
\author{A. Biltmo}
\author{P. Henelius}
\affiliation{Dept. of Theoretical Physics, Royal Institute of Technology, SE-106 91 Stockholm, Sweden}

\date{\today}

\begin{abstract}
Using Monte Carlo simulations we confirm that the rare-earth compound LiHoF$_4$ is a very good realization of a dipolar Ising model. With only one free parameter our calculations for the magnetization, specific heat and inverse susceptibility match experimental data at a quantitative level in the single Kelvin temperature range, including the ferromagnetic transition at 1.53 K. Using parallel tempering methods and reaching system sizes up to 32000 dipoles with periodic boundary conditions we are able to give strong direct evidence of the logarithmic corrections predicted in renormalization group theory. Due to the long range and angular dependence of the dipolar model sample shape and domains play a crucial role in the ordered state. We go beyond Griffiths's theorem and consider surface corrections arising in finite macroscopic samples leading to a theory of magnetic domains. We predict that the ground-state domain structure for cylinders with a demagnetization factor $N>0$ consists of thin parallel sheets of opposite magnetization, with a width depending on the demagnetization factor.
\end{abstract}


\maketitle

The use of effective theories is one of the primary modus operandi of modern physics. Frequently the effective models give only a qualitatively accurate description of the phenomena under investigation, due to corrections that are omitted and free parameters that may be hard to determine experimentally. Finding experimental systems that display striking phenomena, are accurately described by a simple model, and have few or no free parameters, is important since it enables detailed comparison between experiments, theory and numerical simulations. 

The rare earth magnet LiHo$_x$Y$_{1-x}$F$_4$ displays an array of fascinating magnetic phenomena such as quantum phase transitions,\cite{rose96} spin-glass behavior,\cite{rose90} and persistent coherent oscillations \cite{ghos02}. Yet at least the pure material LiHoF$_4$ is believed to be described by one of the most fundamental models in condensed matter physics: the two-state Ising model. Materials such as the antiferromagnets DyPO$_4$ and DyAlG have been shown to be accurately described by a short-ranged Ising model.\cite{wolf00} In LiHoF$_4$, on the other hand, the magnetic properties are dominated by the dipolar interaction. Since the interaction strength is set by the known g factor this makes it possible to determine the effective model to high accuracy. However, the inherent frustration and long-range of the dipolar model make direct numerical simulations demanding. Using a parallel tempering Monte Carlo (MC) method that is essentially free of systematic errors (apart from finite-size effects) we go beyond mean-field theory and explicitly demonstrate that the experimental data for LiHoF$_4$ is indeed in quantitative agreement with the dipolar Ising model.

The magnetic properties of LiHoF$_4$ originate in the $4f$-electrons of the Ho$^{3+}$ ions which sit in a tetragonal lattice with a unit cell of size (1,1,2.077) in units of $a=5.175 \mathring{A}$. 
According to Hund's rules the holmium ion has a $^5I_8$ ground state, but the crystal field partially lifts the 17-fold degeneracy, and the resulting doubly-degenerate ground state is separated from the first excited state by 11 K. This separation of energy levels enables a projection of the full Hamiltonian onto the ground state subspace.\cite{chak04} The matrix elements of the operators $J^x$ and $J^y$ vanish in this subspace, and the effective model is the dipolar Ising model
\begin{equation}
H= \frac{J_d}{2} \sum_{i,j} \frac{r_{ij}^2-3z_{ij}^2}{r_{ij}^5} \sigma_i^z \sigma_j^z + \frac{J_e}{2}\sum_{\langle i j \rangle } \sigma_i^z \sigma_j^z.
\label{eq:Hamiltonian}
\end{equation}
The dipolar coupling constant is given by $J_d=(g\mu_B/2)^2/a^3=0.214 $ K due to the renormalized g factor $\simeq 13.8$, which can be computed from the crystal-field Hamiltonian\cite{chak04}, or deduced from the experimental high-temperature susceptibility\cite{CoJoSiWe75}. The only free parameter in the model, the weak exchange interaction, has no fundamental physical effect on the system other than to alter the critical temperature ($T_c$)\cite{chak04,bilt07}. We set it to $J_{e}=0.12 K$ to reduce the $T_c$ of the model from 1.91 K to the experimental value 1.53 K.

The study of the dipolar interaction has a long and interesting history. Due to demagnetization effects, Luttinger and Tisza \cite{lutt46} found a ground state energy that depends on both lattice structure and sample shape. 
Griffiths later gave a proof that the free energy, without applied fields, is independent of sample shape.\cite{grif68} The apparent contradiction is explained in terms of domain formation, allowing the magnetic order to vary from one macroscopic part of the system to the next. Experimentally this has been demonstrated since measurements of the specific heat for LiHoF$_4$ show no apparent shape dependence,\cite{menn84} and needle shaped domains have been observed close to the transition.\cite{batt75} 

The dipolar interaction has several properties that complicate a numerical treatment of the model. Inherent frustration combined with the long range makes MC equilibration cumbersome at low temperatures, requiring long simulation runs to reach equilibrium. To handle the long range of the interactions we employ the method of Ewald summation\cite{ewal21}. This method not only gives improved numerical convergence due to the use of periodic boundary conditions, but also includes an Ewald parameter which emulates different sample shapes. Our simulations have been carried out using single-flip parallel tempering MC, since cluster methods are of limited use in frustrated systems. 
The MC sample is of size $L^3$ unit cells, with the linear size $L$ ranging from 10 to 20 (4000 to 32000 spins) throughout the study. In most figures we use about 100 temperature points resulting in smooth curves.

\begin{figure}[htp]
  \resizebox{7cm}{!}{\includegraphics[type=eps, ext=.eps,
    read=.eps,clip=true]{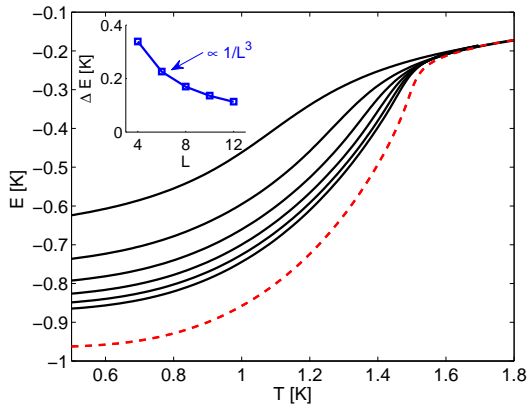}}
  \caption{The internal energy for spherical (black lines) and cylindrical (dashed red line) boundary conditions. For the spherical boundary the MC cell is of size $L^3$ unit cells with $L=10,12,14, 16$ and 18 from top to bottom. The inset shows the difference of the two energies as a function of system size at T=0.5K. }
\label{ene}
\end{figure}

\begin{figure}[htp]
  \resizebox{7cm}{!}{\includegraphics[type=eps, ext=.eps,
    read=.eps,clip=true]{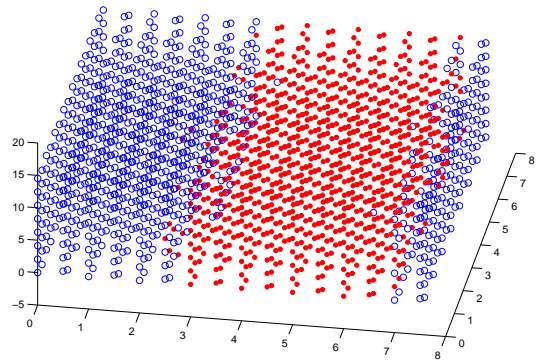}}
  \caption{Typical spin configuration for spherical boundary conditions at $T=0.5T_c$.}
\label{spinconf}
\end{figure}

In order to investigate the convergence to the thermodynamic limit we first consider the 
effects of different sample shapes on the internal energy in Fig.~(\ref{ene}). The calculation is performed for a long needle (with demagnetization factor N=0) and for a sphere (N=4$\pi$/3). The energy for the needle converges quickly in system size and we show the converged curve. The energy for the sphere coincides with the energy for the needle above $T_c$, but shows large finite-size corrections below $T_c$. The needle orders ferromagnetically and according to Griffiths's theorem the infinite spherical sample must have the same energy. If the spherical sample forms ferromagnetic domains that cancel the internal magnetic field, then the two energies will be equal.
 The formation of domains can be seen directly in the simulations of the spherical sample, and in Fig.~(\ref{spinconf}) we show a typical spin configuration at $T=0.5 T_c$. 
In the inset of Fig.~(\ref{ene}) we see that the difference of the two energies decreases as the inverse of the volume of the system.

\begin{figure}[htp]
  \resizebox{7cm}{!}{\includegraphics[type=eps, ext=.eps,read=.eps,clip=true]{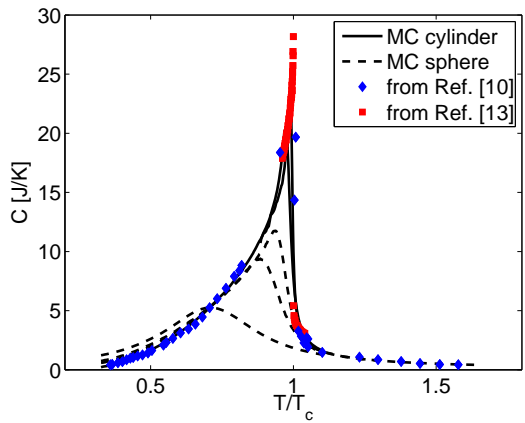}}
  \caption{The specific heat capacity for spherical (L=12,10 and 8 from top to bottom) and cylindrical boundary conditions (L=20 and 18 from top to bottom), and experimental data for a spherical sample \cite{menn84} as well as an oblate sample\cite{NikEll01, GrHuFo80}.}
\label{specheat}
\end{figure}

In order to verify the accuracy of our effective model for LiHoF$_4$ we make detailed comparisons between our calculations and existing experimental data. In Fig.~(\ref{specheat}) we show the specific heat measurements from Refs.~\onlinecite{menn84, GrHuFo80} and our calculations for spherical and cylindrical (N=0) samples. 
Again we find that the numerical results for spherical samples show slow convergence, while the results for zero demagnetization factor have converged, except for close to the critical temperature, where finite-size effects are still visible. In the range of converged data the numerical results agree very well with the experimental data.

\begin{figure}[htp]
  \resizebox{7cm}{!}{\includegraphics[type=eps, ext=.eps,
    read=.eps,clip=true]{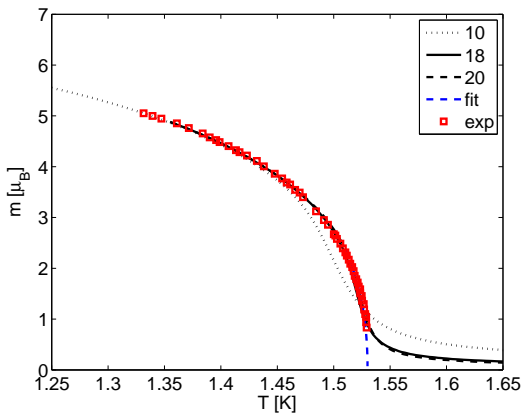}}
  \caption{Magnetization as a function of temperature for system sizes L=10,18 and 20, as well as experimental data\cite{GrHuFo80}. Close to the critical temperature we show a fit to our data including logarithmic corrections.}
\label{mag}
\end{figure}

Next we compare the spontaneous domain magnetization measured by Griffin et al.\cite{GrHuFo80} with our simulations for zero demagnetization factor in Fig.~(\ref{mag}). The experimental data is only determined up to a constant, and we have normalized the experimental data to agree with our calculations at 1.35 K. The agreement is very good all the way up to about $0.96T_c$, where finite size effects become visible in our largest system sizes. Below we will analyze our data more carefully and demonstrate that we can directly observe logarithmic corrections to the mean-field magnetization.

\begin{figure}[htp]
  \resizebox{7cm}{!}{\includegraphics[type=eps, ext=.eps,
    read=.eps,clip=true]{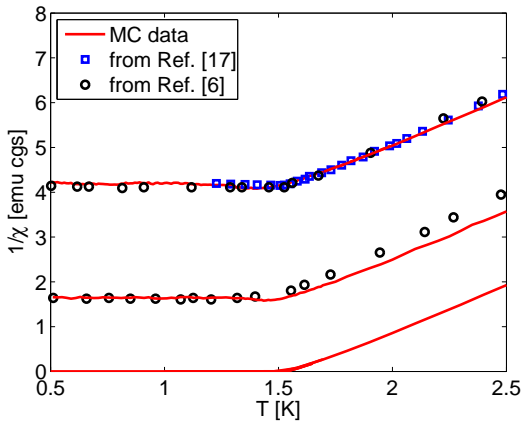}}
  \caption{The inverse magnetic susceptibility $1/\chi$ from experiments and simulations. The three sets of curves correspond to N=$4\pi/3$ (spherical sample), 1.65 and 0 from top to bottom.}
\label{invchi}
\end{figure}

Finally we consider the inverse susceptibility as measured by Cooke et al\cite{CoJoSiWe75} as a function of different demagnetization factors. In Fig.~(\ref{invchi}) we demonstrate that the agreement with experiments again is very good for $N=4\pi/3$ (spherical sample) and N=1.65. The interesting behavior of the the inverse susceptibility below $T_c$ can be understood in terms of the domains.\cite{CoJoSiWe75} With no disorder the walls are free to move and arrange themselves to cancel the internal field: ${\bf H}_{int} = {\bf H}-N{\bf M}=0$, resulting in a constant susceptibility, $\chi={\bf M}/{\bf H}=1/N$, below $T_c$. We find it remarkable that the numerical simulations for limited system sizes have converged so well, even if the domain size is much smaller than in the real material. Above $T_c$ the Curie-Weiss law is followed. For the $N=1.65$ sample we note that the initial slope of the experimental data is slightly higher than the MC data, indicative of a higher g factor. 

\begin{figure}[htp]
  \resizebox{7cm}{!}{\includegraphics[type=eps, ext=.eps,
    read=.eps,clip=true]{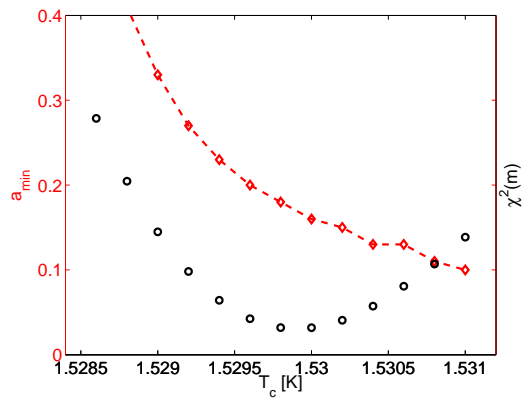}}
  \caption{Exponent $a$ of the logarithmic correction and $\chi^2$ for optimal fit of the magnetization curve as a function of $T_c$. }
\label{RMSElog}
\end{figure}

The upper critical dimension is three for uniaxial dipolar interactions, and according to renormalization group theory, the magnetization, susceptibility and specific heat are predicted to have logarithmic corrections of the form $\log|(T-T_c)/T_0|^{1/3}$, where $T_0$ is an effective temperature.\cite{LarKhm69} Experimentally, logarithmic corrections have been convincingly seen in the magnetization\cite{GrHuFo80} of LiHoF$_4$ and the specific heat of LiTbF$_4$\cite{ahle75}, but not in the susceptibility.\cite{BeReLaWa78} Numerical studies of the dipolar model have applied finite-size scaling to detect the logarithmic corrections,\cite{XuBeRa92,klop06} but since we have results for large system sizes (32000 spins) we directly fit a curve of the form
\begin{equation}
m(t) \sim (T-T_c)^{1/2} |\log|(T-T_c)/T_0||^{a}
\end{equation}
to the part of the critical region where the MC data has converged. For different values of $T_c$ we let the $T_0$ and the exponent $a$ vary and display the value of $a$ that gives the best fit, together with the corresponding $\chi^2$ value in Fig.~(\ref{RMSElog}). There is a minimum in $\chi^2$ around $a=0.18$, giving numerical evidence of non-zero logarithmic corrections. However, the exact value of the optimal $T_c$ and $a$ depends on the temperature interval included in the fit, but a finite value of $a$ does improve the fit.

\begin{figure}[htp]
  \resizebox{7cm}{!}{\includegraphics[type=eps, ext=.eps,
    read=.eps,clip=true]{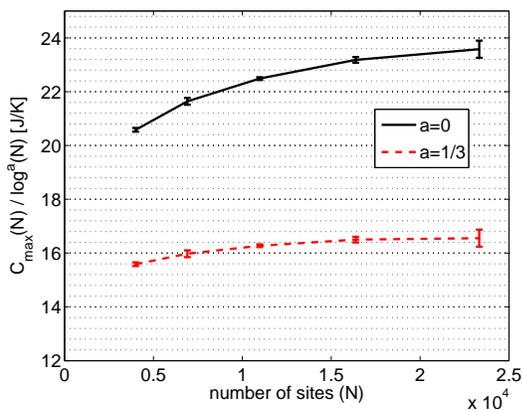}}
  \caption{The peak of the heat capacity grows logarithmically with system size.}
\label{peakClog}
\end{figure}

For the thermodynamic quantities that diverge at the critical point we have not been able to use the same direct fit due to the large finite-size effects. However, convincing evidence for logarithmic corrections in the heat capacity can still be obtained by plotting the peak height against a logarithmically corrected system size. This curve should tend to a constant for large system sizes, and as can be seen in Fig.~(\ref{peakClog}) the curve levels out significantly faster for the predicted exponent $a=1/3$ than for the mean-field result $a=0$.

Griffiths's theorem predicts the formation of domains in order to make the free energy independent of sample shape, but it does not answer the fundamental question of the size and shape of the domains. The domain structure is the result of an energy balance: introducing a domain wall increases the dipolar energy of Eq.~(\ref{eq:Hamiltonian}), whereas the magnetostatic energy\cite{kitt49} is decreased. The energy per spin is of the form $E=C_1/D+ C_2\cdot D + C_3$, where $D$ is the linear size of the domain, $C_1$ the domain wall energy, $C_2$ the magnetostatic energy and $C_3$ is the energy of the ferromagnetic ground state for $N=0$. 

\begin{figure}[htp]
  \resizebox{7cm}{!}{\includegraphics[type=eps, ext=.eps,
    read=.eps,clip=true]{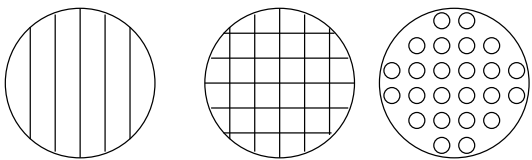}}
  \caption{Domain configurations of parallel sheets ($E_2$), checkerboard pattern ($E_4$) and cylinders ($E_c$).}
\label{configs}
\end{figure}

\begin{figure}[htp]
  \resizebox{7cm}{!}{\includegraphics[type=eps, ext=.eps,
    read=.eps,clip=true]{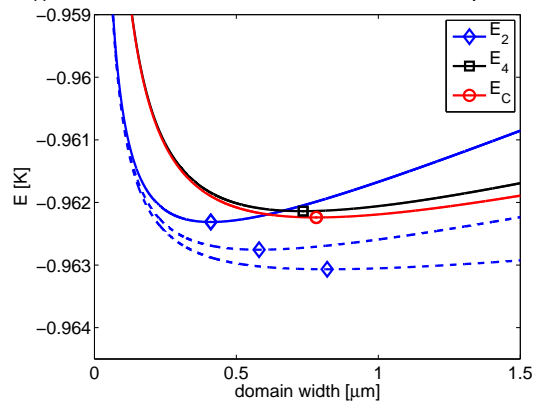}}
  \caption{The energy per spin for parallel sheets ($E_2$), checkerboard ($E_4$) and cylindrical ($E_c$) domain structures for a cylinder of diameter 3.2 mm and height 4.8 mm (solid curves). Dashed curves show $E_2$ for cylinders of increasing heights 9.6 mm and 19.2 mm.}
\label{dom_ene}
\end{figure}

The magnetostatic energy arises from the surface charge of the domain configuration. In the limit of infinite system size this term vanishes, and the energy is given by $E=C_1/D + C_3$, which is the limit considered by Griffiths. We consequently see that as the domain size increases the energy approaches $C_3$. The energy of a particular domain configuration can be calculated as a function of $D$ using the Ewald summation, enabling us to extract the constant $C_1$ for any domain configuration. 

Here we consider three domain structures that are likely candidates for the ground state of LiHoF$_4$: thin parallel sheets ($E_2$), checker board ($E_4$) and cylinders ($E_c$), depicted in Fig.~(\ref{configs}). Kittel has calculated the corresponding magnetostatic surface energy density as 0.85, 0.53 and 0.37 $\mu M^2D$.\cite{kitt49} In Fig.~(\ref{dom_ene}) we plot the energy per spin as a function of domain size in the fully polarized ground state for the three configurations. The configuration of parallel sheets has the lowest energy, and we predict that this is the ground state domain structure. The calculation was done for a cylinder of diameter 3.2 mm and length 4.8 mm. We also show $E_2$ for lengths 9.6 mm and 19.2 mm. As the demagnetization factor of the cylinder decreases, the size of the domains grows, as shown in Fig.~(\ref{dom_ene}), but $E_2$ remains the lowest energy. From Fig.~(\ref{dom_ene}) we see that for finite samples there is actually a small shape dependence of the energy, which disappears in the limit of infinite system size considered by Griffiths. 

We have provided strong evidence that the rare-earth magnet LiHoF$_4$ is a very good realization of the dipolar Ising model. This enables detailed comparisons of theory, experiments and simulations. As examples of this we verify the logarithmic corrections predicted by renormalization group theory for the dipolar model, and we predict that the ground state domain configuration for cylindrically shaped samples consists of thin parallel sheets. Since the domain walls have no width, very clean single crystals with no signs of strain are available, and domains appear naturally in MC simulations, LiHoF$_4$ is an excellent testing ground for theories of domains. In particular we believe there is much further scope for the study of domain-wall motion in the presence of disorder and transverse fields.

This work was supported by the G\"oran Gustafsson foundation and the Swedish Research Council.

\bibliography{artikel3}

\end{document}